\documentclass[twocolumn,superscriptaddress,showpacs,amsfonts]{revtex4-1}
\pdfoutput=1

\usepackage{epsfig}
\usepackage{graphicx}
\usepackage{amsmath}
\usepackage{amssymb}
\usepackage{color}
\usepackage{bm}

\begin{document}

\title{Control of magnetism in singlet-triplet superconducting heterostructures}

\author{Alfonso Romano}
\affiliation{CNR-SPIN, I-84084 Fisciano (Salerno), Italy and
Dipartimento di Fisica ``E. R. Caianiello'', Universit\`a di
Salerno, I-84084 Fisciano (Salerno), Italy}

\author{Paola Gentile}
\affiliation{CNR-SPIN, I-84084 Fisciano (Salerno), Italy and
Dipartimento di Fisica ``E. R. Caianiello'', Universit\`a di
Salerno, I-84084 Fisciano (Salerno), Italy}
\author{Canio Noce}
\affiliation{CNR-SPIN, I-84084 Fisciano (Salerno), Italy and
Dipartimento di Fisica ``E. R. Caianiello'', Universit\`a di
Salerno, I-84084 Fisciano (Salerno), Italy}
\author{Ilya Vekhter}
\affiliation{Department of Physics and Astronomy, Louisiana State
University, Baton Rouge, Louisiana, 70803, USA}
\author{Mario Cuoco}
\affiliation{CNR-SPIN, I-84084 Fisciano (Salerno), Italy and
Dipartimento di Fisica ``E. R. Caianiello'', Universit\`a di
Salerno, I-84084 Fisciano (Salerno), Italy}

\begin{abstract}
We analyze the magnetization at the interface between singlet and triplet superconductors and show that its direction and
dependence on the phase difference across the junction are strongly tied to the structure of the triplet order parameter
as well as to the pairing interactions.
We consider equal spin helical, opposite spin chiral, and mixed symmetry pairing on the triplet side
and show that the magnetization vanishes at $\phi=0$ only in the first case, follows approximately a
$\cos\phi$ behavior for the second, and shows higher harmonics for the last configuration.
We trace the origin of the magnetization to the magnetic structure of the Andreev bound states near the interface,
and provide a symmetry-based explanation of the results.
Our findings can be used to control the magnetization in superconducting heterostructures and
to test symmetries of spin-triplet superconductors.
\end{abstract}


\maketitle


{\it Introduction.} Attempts to combine dissipationless transport in superconductors with the
control of spin currents  has driven many recent
studies of superconducting heterostructures~\cite{Robinson2015}. In the vast majority of superconductors the conduction
electrons pair in a spin-singlet state, freezing out the spin degrees of freedom at low temperature.
However, triplet correlations near interfaces with magnetic materials produce equal-spin Cooper
pairs that can propagate through a magnet leading, for example,  to the long-range proximity effect~\cite{Bergeret2005,Buzdin2005,Tanaka2007}.

The alternative pathway to spin control by superconductivity is via utilizing compounds that support spin-triplet pairing.
The number of known triplet superconductors (TSCs) has been growing steadily, and now includes UPt$_3$~\cite{Stewart1984},
ferromagnetic superconductors such as UGe$_2$, URhGe, UIr and UCoGe~\cite{Saxena2000,Aoki2001,Akazawa2004},
quasi one-dimensional organic system (TMTSF)$_2$X (X=ClO$_4$ and PF$_6$)~\cite{Lebed1986,Lee1997}.
Singlet and triplet states are mixed if the material lacks inversion symmetry~\cite{Frigeri2004,Fujimoto2007},
and among the non-centrosymmetric superconductors Li$_2$Pt$_3$B has the clearest indication of a
significant triplet component~\cite{Yuan2006,Nishiyama2007}.
The strongest evidence for triplet superconductivity has emerged for Sr$_2$RuO$_4$~\cite{Maeno2012,MaenoRMP}. 
Existence of very pure single crystals with the perovskite structure made this material a testbed for studying heterostructures based on triplet superconductivity~\cite{Anwar2014,Gentile2013}.

From energetic considerations most of the non-magnetic superconducting compounds that support triplet pairing should have a
unitary order parameter~\cite{Sigrist1996}, so that the Cooper pairs do not have a net average spin-derived magnetic moment~\cite{Ueda1991}.
Hence, triplet superconductors cannot by default be assumed to support dissipationless spin transport. However, in conjunction with superconducting
orders of different symmetry, nontrivial spin aspects of triplet superconductivity appear.
Andreev bound states in singlet-triplet Josephson heterostructures were analyzed~\cite{Asano2003,Yip1993,Yoshida1999,Asano2005}, and
spin-accumulation was found  when a phase difference is established across the junction~\cite{Kwon2004,Sengupta2008,Lu09}.
In parallel, it was shown that admixture of the subdominant order may lead to spin accumulation, spin and charge currents near a boundary of a chiral triplet superconductor~\cite{Romano2013}
or anomalous flux response in mesoscopic loops~\cite{Zha2015}.

In this Letter we show that the magnetism in singlet-triplet superconducting heterostructures is fundamentally linked to the nature of the triplet pairing.
We consider a microscopic model for a high-transparency interface between a singlet and a unitary triplet (or mixed parity, see below)
superconductor, and self-consistently solve the corresponding Bogoliubov-de Gennes (BdG) equations to obtain the energy spectrum.
In all cases we find spin-splitting of the Andreev bound states (ABS) near the interface, leading to a magnetization parallel to the spin-triplet $\bm d$-vector.
However, the variation of the magnetization $M$ with the phase difference $\phi$ across the junction is non-trivial, and
depends on the nature of the pairing interaction on the triplet side. Our main results are summarized in Fig.~1, where we consider three distinct cases.
If there is no singlet component of the pairing interaction on the triplet side, $M\propto \sin \phi$ only appears under a finite phase difference,
in agreement with Refs.~\onlinecite{Kwon2004,Sengupta2008}. This is realized, for example, for equal spin pairing states. Real space interactions leading to triplet
$S^z=0$ pairing often promote a (subdominant) singlet pairing, so that the singlet
amplitude persists into TSC near the interface. The magnetization resulting from this mixed symmetry, $M\propto \cos\phi$,  already occurs at $\phi=0$.
Finally, motivated by the results of Ref.~\onlinecite{Romano2013},
we consider a mixed parity superconductor in contact with a singlet counterpart, and
show that the magnetization has a complex dependence on $\phi$.

These results show that the phase difference can be used to
control interface magnetization. In addition, the interface magnetization probes the spin structure of the pairing interactions.
This finding is especially relevant for Sr$_2$RuO$_4$.
There is no consensus on the exact form of the triplet pairing in this system, and NMR Knight shift measurements \cite{Murakawa2004},
absence of edge currents \cite{Kallin2012}, and the observation of half-quantum vortices \cite{Jang2011} put in doubt strong
pinning of the triplet order spin vector, $\bm d (\bm k)$, to the crystalline $c$-axis.
The chiral,
$\bm d (\bm k)=\widehat{\bm z} (k_x+ik_y)$, and the helical, $\bm d (\bm k)=\widehat{\bm x} k_y+ \widehat{\bm y}k_x$, states compete closely, ~\cite{Scaffidi2014} 
and distinguishing between these two possibilities with similar bulk gap is an important application of our results.

%
\begin{figure}
\begin{center}
\includegraphics[width=0.4\textwidth]{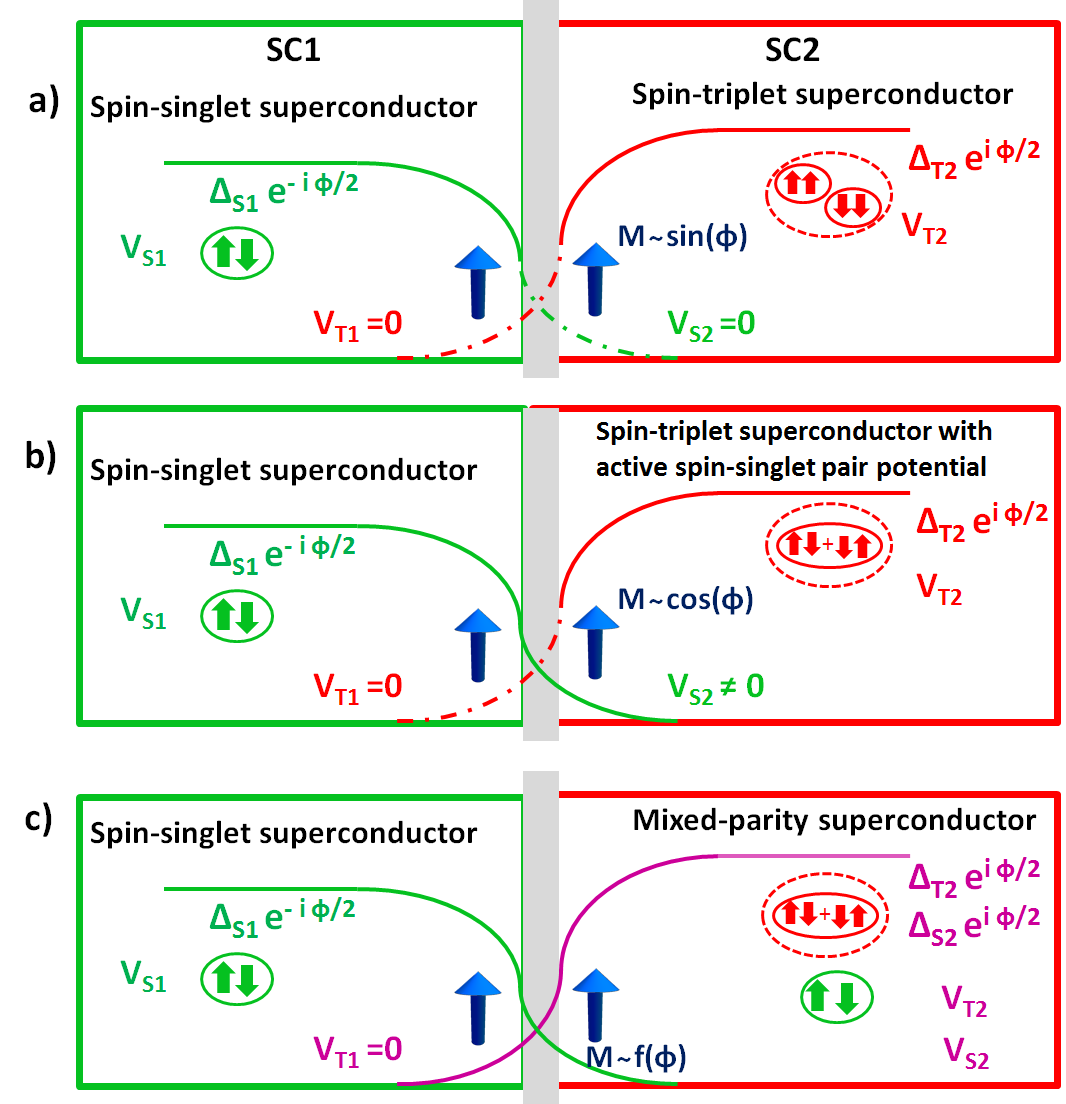}
\end{center}
\caption{Schematics for the superconducting heterostructures between spin-singlet (SC1) and triplet-active (SC2) superconductors. Panel a): helical equal spin pairing state with no pairing interaction in the singlet channel. Panel b):  chiral pairing with finite interaction in the singlet channel (see discussion in text). Panel c): mixed-parity order parameter near the interface. $V_{Si}$ ($V_{Ti}$) are the couplings in the singlet (triplet) channel for sides $i=1,2$. We show the profile of the order parameters (full line),
with dot-dashed line for the proximity-induced order. We also indicate the qualitative dependence of the interface magnetization on the phase $\phi$. }
\label{fig1}
\end{figure}
%

{\it Model and formalism.} We consider a two-dimensional
lateral heterostructure made of two superconductors, SC1 and SC2, having spin-singlet (SC1)
and spin-triplet or mixed-parity (SC2) pairing.
The $x$ and
$y$ planar directions are perpendicular and parallel to the
SC1/SC2 interface, respectively. The system is uniform along the $y$
axis, so that the translational symmetry is broken only in the $x$
direction.
The Hamiltonian is then defined on a square lattice of
size $L\times L$ (the lattice constant is unity), with periodic
boundary conditions along $y$,
\begin{eqnarray}
H=H_1+H_2+H_{12}
\label{eq:Htot}
\end{eqnarray}
with $H_m$ ($m=1,2$) being the Hamiltonians of each superconducting side
\begin{eqnarray} &&H_m=
-t_{m}\sum_{\langle \mathbf{i} ,\mathbf{j} \rangle,\,\sigma}
(c^{\dagger}_{\mathbf{i}\,\sigma}
c_{\mathbf{j}\,\sigma}+\text{h.c.}) -\mu_{m} \sum_{\mathbf{i},\sigma}
n_{\mathbf{i}\sigma}  \\
&&- \sum_{\langle \mathbf{i} ,\mathbf{j} \rangle} V^{\sigma\sigma'}_{m} n_{\mathbf{i} \sigma}
n_{\mathbf{j}\sigma^{'}} - U{_{m}}\sum_{\mathbf{i}} n_{\mathbf{i}\uparrow} n_{\mathbf{i}\downarrow} \,,  \nonumber
\label{eq:Hm}
\end{eqnarray}
and the interface term,
\begin{eqnarray}
H_{12}=\sum_{\mathbf{\delta=\pm 1}} t_{\perp} (c^{\dagger}_{\mathbf{0}\,\sigma}
c_{\mathbf{\delta}\,\sigma}+\text{h.c.}) \; .
\label{eq:Hint}
\end{eqnarray}

Here the lattice sites are labelled by
$\mathbf{i}\equiv\{i_x,i_y\}$, with $i_x$ and $i_y$ integers between
$-L/2$ and $L/2$, $\langle \mathbf{i},\mathbf{j}\rangle $ denote
nearest-neighbor sites, and $\mu$ is the chemical potential.
Labels $\mathbf{0}=\{0,i_y\}$ and $\mathbf{\pm 1}=\{\pm 1,i_y\}$ denote the sites at the interface and their nearest-neighbors, respectively.
The attractive interaction $-V^{\sigma\sigma'}_{m}$ $(V_m>0)$ can be chosen to be effective
in the $S_z=1,0,-1$ projections for the the TSC and/or in the $\bm S=0$ singlet sector.
The local attractive term $-U_m$ ($U>0$) only promotes spin-singlet pairing. For simplicity we take
$t_1=t_2=t_{\perp}=t$, and use the hopping parameter $t$ as a unit of energy. Below, in the description of the results
we discuss the qualitative consequences of relaxing this assumption.

To investigate the model of Eq.~\eqref{eq:Htot} we decouple the interaction term in
the Hartree-Fock approximation by introducing the pairing
amplitude on a bond, $\Delta_{\mathbf{i}\mathbf{j}}^{\sigma \sigma'}=\langle
c_{\mathbf{i}\,\sigma} c_{\mathbf{j}\,\sigma'} \rangle$, and on-site $\Delta_{0}=\langle
c_{\mathbf{i}\,\uparrow} c_{\mathbf{i}\,\downarrow} \rangle$, so that
\begin{equation}
V^{\sigma \sigma'} n_{\mathbf{i}\sigma} n_{\mathbf{j}\sigma'} \simeq
V^{\sigma \sigma'} (\Delta^{\sigma \sigma'}_{\mathbf{i}\mathbf{j}} \,
c^{\dagger}_{\mathbf{j}\,\sigma'}
c^{\dagger}_{\mathbf{i}\,\sigma}+\bar{\Delta}^{\sigma \sigma'}_{\mathbf{i}\mathbf{j}}
c_{\mathbf{i}\,\sigma}
c_{\mathbf{j}\,\sigma'}-|\Delta^{\sigma \sigma'}_{\mathbf{i}\mathbf{j}}|^2)\,.
\nonumber
\end{equation}
These expressions yield the spin singlet ($S$) and triplet
($T$) components in the $S_z=0$ sector,
$\Delta^{S,T}=(\Delta_{\mathbf{i}\mathbf{j}}^{\uparrow \downarrow}\pm\Delta_{\mathbf{j}\mathbf{i}}^{\uparrow \downarrow})/2$, and the
triplet pairing in the $S_z=\{1,-1\}$ sectors, $\Delta^{\sigma \sigma}_{T}=\Delta_{\mathbf{i}\mathbf{j}}^{\sigma \sigma}$.
They define in turn the superconducting pair amplitudes with $s$-
or $p$-wave symmetry, i.e.
$\Delta_{s}(\mathbf{i})=(\Delta^{S}_{\mathbf{i},\mathbf{i}+\text{\^{x}}}+\Delta^{S}_{\mathbf{i},\mathbf{i}-\text{\^{x}}}+
\Delta^{S}_{\mathbf{i},\mathbf{i}+\text{\^{y}}}+\Delta^{S}_{\mathbf{i},\mathbf{i}-\text{\^{y}}})/4$,
$\Delta_{p_{x(y)}}(\mathbf{i})=(\Delta^{T}_{\mathbf{i},\mathbf{i}+\text{\^{x}}(\text{\^{y}})}-\Delta^{T}_{\mathbf{i},\mathbf{i}-\text{\^{x}}(\text{\^{y}})})/2$ and
$\Delta_{p_{x(y)}}^{\sigma \sigma}(\mathbf{i})=\Delta^{\sigma\sigma}_{\mathbf{i},\mathbf{i}+\text{\^{x}}(\text{\^{y}})}$,
which are then determined self-consistently~\cite{Cuoco08}.
It is important to note that, while the terms $V^{\uparrow\uparrow}$,$V^{\downarrow\downarrow}$
generate equal spin triplet pairing, the coupling $V^{\uparrow\downarrow}$ generically promotes both triplet $S^z=0$ and non-local singlet (extended $s$-wave) pairing. Similar decoupling of the $U$ term produces local singlet pairing only.

By suitable choices of the interactions $V^{\sigma\sigma'}$ and $U$ we can therefore model
interfaces between superconductors with different symmetries of the order parameter.
We take the most conventional SC1 with $U\neq 0, V^{\sigma\sigma'}=0$, and compare the results for three different choices of SC2.
Case a) is equal spin triplet pairing, $V^{\uparrow\uparrow}$=$V^{\downarrow\downarrow} \neq 0$; $V^{\uparrow\downarrow}=U=0$.
In the bulk this corresponds to the helical state, $\widehat{\bm\Delta} (\bm k)=i(\bm d(\bm k)\cdot\bm\sigma)\sigma^y$ with $\bm d (\bm k)=\widehat{\bm x} k_y+ \widehat{\bm y}k_x$,
where $\widehat{\bm\Delta} (\bm k)=V^{-1/2}\int d\bm r_{\bf ij}\widehat{\bm\Delta}_{\bf ij}\exp(i\bm k\cdot\bm r_{\bf ij})$ and $\widehat{\bm\Delta}_{\bf ij}$
is the order parameter in spin space defined above. Case b) is a TSC with $S^z=0$ and a subdominant extended $s$-wave pairing, $V^{\uparrow\uparrow}$=$V^{\downarrow\downarrow} =U=0$; $V^{\uparrow\downarrow}\neq 0$. Finally, case c) is a TSC with $S^z=0$, $V^{\uparrow\downarrow}\neq 0$, but possible local $s$-wave pairing, $U\neq 0$, while $V^{\uparrow\uparrow}$=$V^{\downarrow\downarrow}=0$.

For the last two situations the most favorable pairing state due to the $V^{\uparrow\downarrow}$ depends on
the electron density, $n$, and  the chiral $k_x+ik_y$ order is
stabilized in the region between low doping $\mu\simeq 1.2$ and
high (low) density ($|\mu|\simeq 2.25)$~\cite{Kuboki01}. Hence we
choose $|\mu|\simeq 1.8$.
All the numerical results
below have been obtained for non-vanishing components of the pairing interaction $V=2.5$, and $U=2.5$, and system size $L=120$. Greater
values of $L$ and modification of the couplings leave the results qualitatively unchanged.
To investigate the effects of the phase difference between the two superconductors, as is commonly done for the Josephson junctions,
we transform the pairing wave-function in the SC1 and SC2 by the phase factors $\exp[-i \phi/2]$ and $\exp[i \phi/2]$, respectively.
Self-consistent solution of the BdG equations gives the spin-resolved energy spectrum of the system, including both bulk and the surface
(Andreev) states. The spectrum determines the surface magnetization, and we compute $M$ by summing the spin expectation values over all the occupied states at zero temperature.

%
\begin{figure}
\begin{center}
\includegraphics[width=0.45\textwidth]{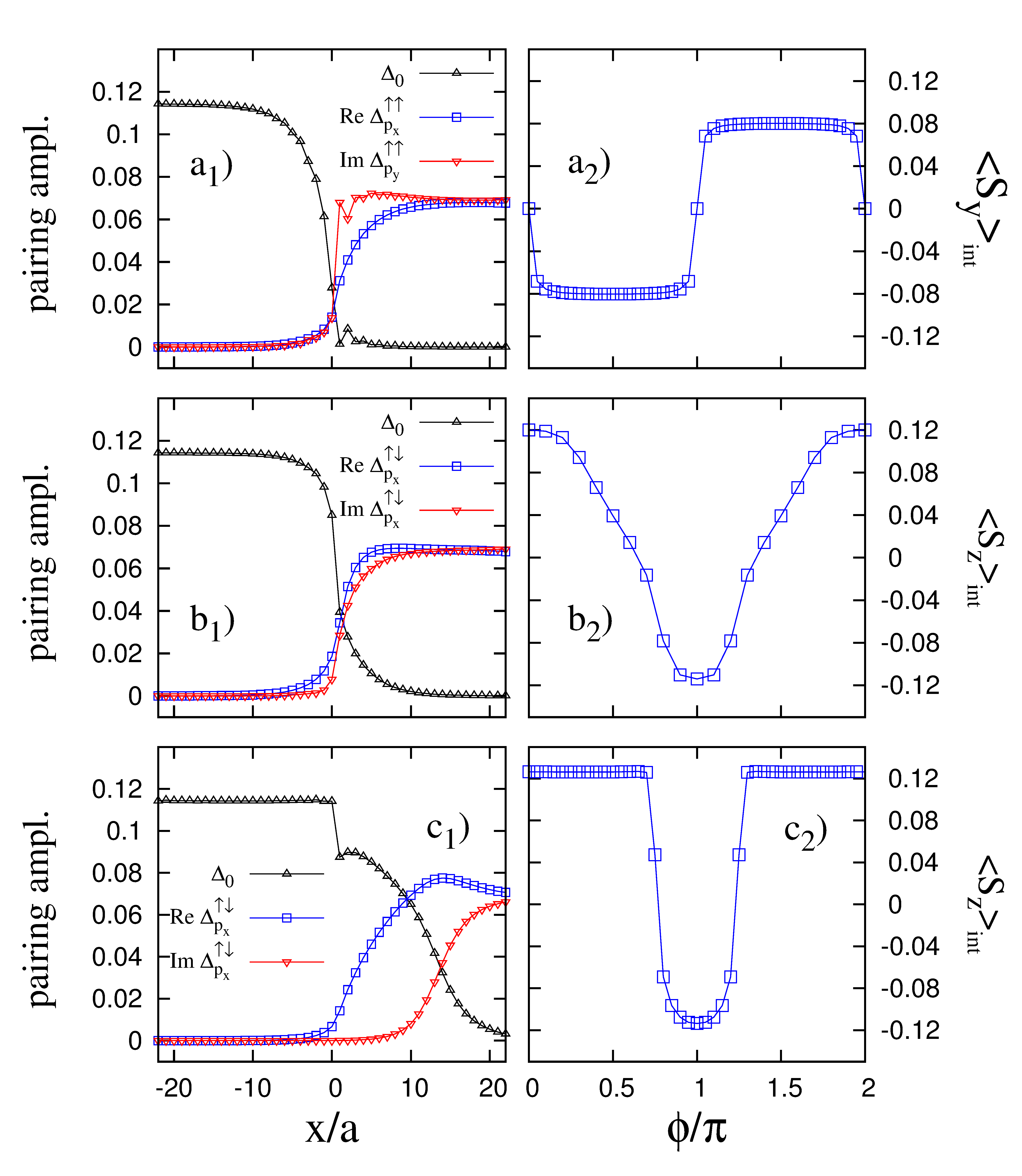}
\end{center}
\caption{Order parameter and magnetization for heterostructures between a singlet (left side of the junction) and
different triplet systems. Panels (a) are for helical equal spin pairing, (b) for chiral $S^z=0$pairing, and (c) for the mixed symmetry case.
Left column shows the evolution of the singlet and triplet pairing amplitudes, while the right column shows the net magnetization along the
direction of the $\bm d$-vector component for which the surface is pairbreaking as a function of the phase difference, $\phi$. }\label{fig2}
\end{figure}
%

{\it Results and discussion.} The interface breaks the translational symmetry in the $x$ direction, and is therefore pair-breaking for the $k_x$ component of the triplet pairing. The differences between the three investigated cases stem from the consequences of this suppression for the interface magnetization. Case a) corresponds to the usual proximity coupling with step-like change of the symmetry of the pairing interaction across the junction, where the reduction of the amplitude of the $p_x$ component simply changes the effective coupling between the two sides, see Figs. 2(a). The magnetism of the Andreev states localized at the interface is due to the splitting of the energy levels for opposite spins.  This splitting appears because the phase shift between the singlet and the triplet enters the continuity condition for the wave function at the boundary with opposite signs for the two spin orientations. Hence the magnetization vanishes at $\phi=0$, and is nearly constant away from this point, see Fig. 2(a2).  The situation is analogous to that discussed in Ref.~\onlinecite{Sengupta2008}, where a step-like change was found without the self-consistency on the order parameters, and assuming a delta-function barrier at the interface (BTK approximation).

The origin of this dependence  is easy to understand from the Ginzburg-Landau (GL) expansion of the free energy density, with the relevant terms $f=am^2+i\,b\,m(\psi^\star\eta_x-\psi\eta_x^\star)$. Here $a>0$ is the inverse susceptibility, $\psi$ ($\eta_x$) are the amplitudes of the singlet (triplet $k_x$) order, and $m$ is the component of the magnetization along the $\bm d$-vector coupled to $k_x$, in our case $\widehat{\bm y}$.  Near the interface where both $\psi=|\psi|e^{-i\phi/2}$ and $\eta_x=|\eta_x|e^{i\phi/2}$ coexist, linear coupling to the magnetization ensures that the mimimum of the free energy is at $m\propto \sin\phi$. The GL analysis is valid near $T_c$ where indeed a $\sin\phi$-like shape develops from the step-like behavior of the same symmetry. \cite{Sengupta2008} The spatial profile of the magnetization, shown in Fig.~3(a), confirms that it is constrained to the region of the order of the coherence length around the junction's interface.

In general the magnetization linearly couples to the component of the $\bm d$-vector for which the surface is pair breaking, and hence will be along the $\bm{\widehat z}$ axis for the remaining two situations. The phase dependence also becomes more complex if there is pairing in the subdominant $s$-wave channel on the triplet side, as in cases b) and c). Recall that we have a chiral triplet order stabilized in the bulk. Reduction in the triplet $k_x$ component near the interface generically implies $\Delta^{\uparrow\downarrow}_{\bf i, i+\widehat{\bf x}}\neq -\Delta^{\uparrow\downarrow}_{\bf i, i-\widehat{\bf x}}$, and allows for admixture of singlet pairing. The difference from the well-known case of subdominant pairing near the surfaces in $d$-wave superconductors~\cite{Fogelstrom1997} is that it is the same coupling constant $V^{\uparrow\downarrow}$ that promotes coupling in both channels, and therefore the admixture is at least parametrically stronger than in the case of subdominant coupling. Indeed, it was found that even small variations in the surface barrier at the boundary of a triplet superconductor generated a substantial admixture of a singlet component and emergence of mixed parity superconductivity near the edge~\cite{Romano2013}. Assuming that it is the subdominant extended $s$-wave that is coupled to the triplet component near the interface, the linear coupling terms in the GL expansion take the form $f_m=b^\prime m[\psi(x)^{*} \partial_{x} \bm{\eta}(x)+\psi(x) \partial_{x} \bm{\eta}(x)^{*}]$, and promote $m\propto\cos\phi$ phase dependence of the magnetization (a similar term was found in the microscopic theory \cite{Kuboki:11}). This is illustrated in Figs. 2(b$_1$) and 2(c$_1$), where the $k_y$ component of the chiral order is suppressed concomitantly with the $k_x$ component (in contrast to Fig. 2(a$_1$)), and the magnetization already exists without the phase difference across the junction.

In principle both $\sin\phi$ and $\cos\phi$ contributions should be present in such a junction, leading to a non-trivial phase dependence. However, we believe that in case b) the competition with the subdominant pairing on the SC2 side causes the triplet components to decay more rapidly towards the interface and save energy via singlet pairing (compare Fig. 2(a$_1$) and Fig. 2(c$_1$)), enhancing the latter term, and exhibiting dominant $\cos\phi$ behavior. Indirectly this is also supported by the magnetization that is more sharply localized near the interface, compare Fig. 3(a) and Fig. 3(b).

Finally, inclusion of the on-site $s$-wave pairing for case c) dramatically extends the range of coexistence of triplet and singlet pairing in SC2, see Fig.~2(c$_1$). This implies a more pronounced competition between the non-gradient and gradient couplings of the superconducting orders to the magnetization, but there is an additional complexity because the ``local'' $s$-wave component suppresses the $k_y$ component of the pairing faster than the $k_x$ component, Fig.~2(c$_1$). The induced $s$-wave order on the SC2 side is phase-locked to the triplet component, yielding a finite magnetization that only weakly depends on the phase for $\phi\in(0,\pi/2)$, but changes sign as the phase difference approaches $\pi$. Presumably this occurs because of additional states appearing due to the singlet component changing sign across the junction. This view is supported by considering the spatial profile of the magnetization, Fig. 3(c), which shows a two-peak structure for small $\phi$, including the contribution away from the interface, but only exhibits a single peak at the interface for $\phi=\pi$. The resulting phase dependence of the magnetization contains many Fourier components.

Analysis of the energy spectrum, Fig. 3(d), supports these conclusions. If the structure of the order parameter near interface plays the major role in the emergence of the magnetization, nearly grazing trajectories must have a significant contribution. Indeed, the spin-split branches of Andreev states crossing the Fermi surface are moved to large values of $k_y/\pi\approx 0.5$. In addition, there is a secondary branch for the occupied states just below the gap edge at large $k_y$, and the peak in the energy dispersion of this state implies a van-Hove-like singularity contributing to the net $M$.

%
\begin{figure}
\begin{center}
\includegraphics[width=0.45\textwidth]{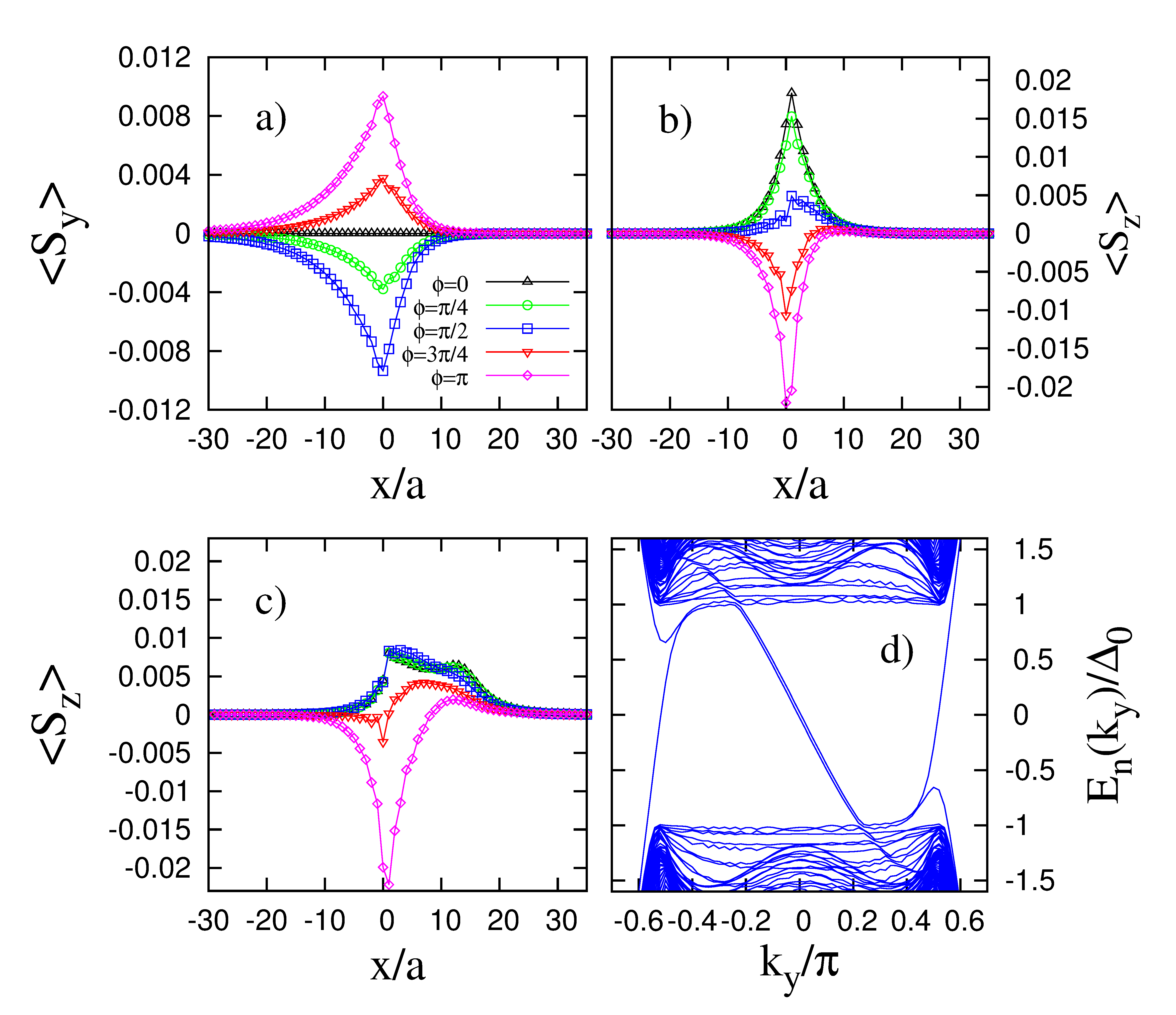}
\end{center}
\caption{Spatial dependence of the spin-polarization. Panels (a)-(c) correspond to the cases of helical,
chiral, and mixed symmetry order parameters as discussed in the text. Panel (d) shows a typical energy spectrum (for mixed symmetry case and $\phi=0$). The thick line through the gap
is the chiral edge state away from the interface. A significant contribution to the magnetization is due to the Andreev state just within the gap at large values of $k_y$ .}\label{fig3}
\end{figure}
%

{\it Conlcusions.} We showed that the magnetic properties of singlet-triplet heterostructures are extraordinary
sensitive to the type of triplet pairing, and to subdominant pairing channels. The dependence
of the magnetization on the phase distinguishes between helical and chiral pairing. Vice versa,
our results allow control of the interface magnetization by the phase difference. We considered an interface of high transparency and identical band
structure on both sides. Relaxing this assumption will reduce the amplitude of the leaking superconducting component. In the first two cases we expect a reduction in the amplitude of $M(\phi)$,
while in the last example the magnetization is weakly sensitive to the interface mismatch except near $\phi\sim\pi$. 
Finally, our results suggest that such junctions act as possible spin pumps.
A voltage bias across the heterostructure would make the phase difference time-dependent,
$\phi=2 e V t$, and the dynamics of the magnetization is determined by this driving force as well as damping.
If damping is small, for our last case of mixed-parity superconductor
in contact with a conventional $s$-wave system, the
magnetization oscillations do not average to zero over a period of $\pi/e V$. Detailed investigation of this effect is left for future work.

I.V. acknowledges support from NSF Grant DMR-1105339.
M.C. and P.G. acknowledge support from FP7- EC programme under FET-
Open grant number 618083 (CNTQC).

\end{document}